\documentclass[%
reprint,
superscriptaddress,
amsmath,amssymb,
aps,
prapplied,
]{revtex4-1}

\usepackage{graphicx}
\usepackage{dcolumn}
\usepackage{bm}
\usepackage{float}
\usepackage{hyperref}

\begin{document}

\preprint{APS/123-QED}

\title[Manuscript for Phys. Rev. Appl.]{A Nanoscale Room-Temperature Multilayer Skyrmionic Synapse \\for Deep Spiking Neural Networks}

\author{Runze Chen}
\thanks{These authors contributed equally to this work}
\affiliation{ 
Nano Engineering and Spintronic Technologies (NEST) Group, Department of Computer Science, School of Engineering, the University of Manchester, Manchester M13 9PL, United Kingdom}

\author{Chen Li}
\thanks{These authors contributed equally to this work}
\affiliation{ 
Advanced Processor Technologies (APT) Group, Department of Computer Science, School of Engineering, the University of Manchester, Manchester M13 9PL, United Kingdom}

\author{Yu Li}
\affiliation{ 
Nano Engineering and Spintronic Technologies (NEST) Group, Department of Computer Science, School of Engineering, the University of Manchester, Manchester M13 9PL, United Kingdom} 

\author{James J. Miles}
\affiliation{ 
Nano Engineering and Spintronic Technologies (NEST) Group, Department of Computer Science, School of Engineering, the University of Manchester, Manchester M13 9PL, United Kingdom}

\author{Giacomo Indiveri}
\affiliation{ 
Institute of Neuroinformatics, University of Zurich and ETH Zurich, 8057 Zurich, Switzerland} 

\author{Steve Furber}
\affiliation{ 
Advanced Processor Technologies (APT) Group, Department of Computer Science, School of Engineering, the University of Manchester, Manchester M13 9PL, United Kingdom} 

\author{Vasilis F. Pavlidis}
\affiliation{ 
Advanced Processor Technologies (APT) Group, Department of Computer Science, School of Engineering, the University of Manchester, Manchester M13 9PL, United Kingdom}

\author{Christoforos Moutafis}
\email{christoforos.moutafis@manchester.ac.uk}
\affiliation{ 
Nano Engineering and Spintronic Technologies (NEST) Group, Department of Computer Science, School of Engineering, the University of Manchester, Manchester M13 9PL, United Kingdom}


\date{\today}

\begin{abstract}
Magnetic skyrmions have attracted considerable interest, especially after their recent experimental demonstration at room temperature in multilayers.
The robustness, nanoscale size and non-volatility of skyrmions have triggered a substantial amount of research on skyrmion-based low-power, ultra-dense nanocomputing and neuromorphic systems such as artificial synapses.
Room-temperature operation is required to integrate skyrmionic synapses in practical future devices.
Here, we numerically propose a nanoscale skyrmionic synapse composed of magnetic multilayers that enables room-temperature device operation tailored for optimal synaptic resolution. 
We demonstrate that when embedding such multilayer skyrmionic synapses in a simple spiking neural network (SNN) with unsupervised learning via the spike-timing-dependent plasticity rule, we can achieve only a $\sim 78\%$ classification accuracy in the MNIST handwritten data set under realistic conditions.
We propose that this performance can be significantly improved to  $\sim 98.61\%$ by using a deep SNN with supervised learning. 
Our results illustrate that the proposed skyrmionic synapse can be a potential candidate for future energy-efficient neuromorphic edge computing.
\end{abstract}

\keywords{Nanoscale skyrmionic synapse, spiking neural networks, neuromorphic computing}
\maketitle

\section{Introduction}
Neuromorphic computing draws inspiration from how the human brain performs extremely energy-efficient computations \cite{mead1990neuromorphic,chicca2014neuromorphic}.
Building ultra-low power cognitive computing systems by mimicking neuro-biological architectures is a promising way to achieve such efficiency.
One of the candidate technologies is spintronics \cite{vzutic2004spintronics,torrejon2017neuromorphic,grollier2020neuromorphic}.
Recently, skyrmion-electronics (`skyrmionics'), a branch of spintronics, has been proposed as a promising building block for next-generation data storage and processing applications \cite{fert2013skyrmions,sampaio2013nucleation}.

\begin{figure*}
\includegraphics[width=172mm]{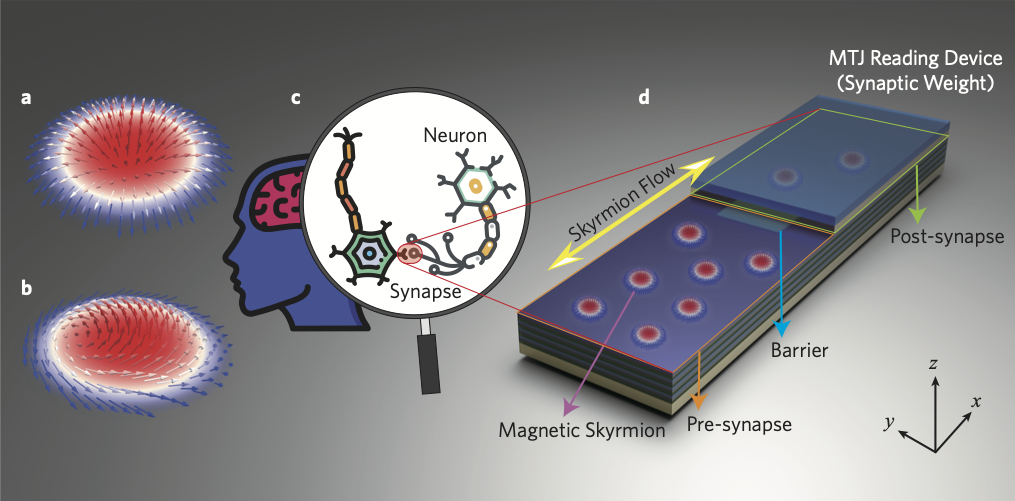}
\caption{\label{fig:1}The proposed skyrmionic synaptic device.
 Illustrations of (a) a N\'{e}el skyrmion (used in this device) and (b) a Bloch skyrmion spin texture. (c) Schematic of biological neurons connected with a synapse. (d) The proposed nanoscale multilayer skyrmionic synapse device based on skyrmion flow between a pre-synapse and a post-synapse region \cite{huang2017magnetic}. The multilayer structure here enables room-temperature operations.}
\end{figure*}

Magnetic skyrmions [Fig. \ref{fig:1}(a, b)] are topologically protected spin textures exhibiting particle-like properties \cite{fert2013skyrmions}.
Skyrmions were recently demonstrated in both bulk non-centrosymmetric chiral magnets and magnetic multilayer (MML) thin films with the existence of Dzyaloshinskii-Moriya interaction (DMI) originating from strong spin-orbit coupling (SOC) and broken inversion symmetry \cite{dzyaloshinsky1958thermodynamic}.
Skyrmion-based computational and storage devices can be envisaged as hybrid solutions with traditional Complementary Metal Oxide Semiconductor (CMOS) technologies that enhance functionality due to their robustness, nanoscale size, and non-volatility \cite{fert2013skyrmions}.
Magnetic skyrmions have recently been demonstrated experimentally at room temperature (RT) for the first time in tailored technologically relevant MMLs \cite{moreau2016additive,boulle2016room} with a diameter of individual skyrmions in the sub-$100$ $\mathrm{nm}$ range \cite{legrand2017room}, which opens the way for their use in future nanocomputing applications.
Magnetic skyrmions can be nucleated, manipulated and deleted via various external stimuli \cite{zhang2020skyrmion,everschor2018perspective}, such as spin-polarised current \cite{fert2013skyrmions,sampaio2013nucleation,legrand2017room}, magnetic field gradient \cite{lai2017improved} and localized heating with laser irradiation \cite{finazzi2013laser,fujita2017ultrafast}.

The enhanced stability and robustness of skyrmions results from their topology \cite{fert2013skyrmions} and energy contributions \cite{rohart2016path}.
Skyrmionic spin textures can be described by the topological charge or skyrmion number \cite{moutafis2009dynamics}, which counts how many times the vector field configuration wraps around a unit sphere and which is therefore an integer. This is defined as:
\begin{equation}
    N = \frac{1}{4\pi} \int \textbf{\textit{m}}\cdot(\partial_x\textbf{\textit{m}}\times\partial_y\textbf{\textit{m}})\textit{dxdy},
\end{equation}
where $N = \pm 1$ for the case of skyrmions and its sign reflects the polarity.
It is possible to have an infinite set of skyrmion solutions with different value and sign of $N$ in chiral magnets \cite{rybakov2019chiral} as well as topologically trivial states with $N = 0$ \cite{moutafis2009dynamics,sampaio2013nucleation}. 

Magnetic skyrmions have attracted considerable interest as information carriers in nanodevices that can emulate biological synapses due to their unique physical characteristics \cite{huang2017magnetic,li2018emerging}.
Micromagnetic simulations of such devices suggest that these devices consume far less energy than conventional chips and other non von Neumann architectures \cite{huang2017magnetic,li2018emerging}.
Meanwhile, a multi-bit skyrmion-based non-volatile storage device has been proposed and numerically simulated, where the synaptic states of the device are modulated by electric pulses shifting the position of skyrmions within the device \cite{bhattacharya2019low}.
More recently, a considerably larger, micrometer-scale, room-temperature ferrimagnetic artificial synapse has been recently experimentally demonstrated \cite{song2020skyrmion}. 
Further possibilities have been proposed to utilize magnetic skyrmions in Artificial Neural Networks (ANNs), Spiking Neural Networks (SNNs) and reservoir computing \cite{ananthanarayanan2009cat,prychynenko2018magnetic}.

Although skyrmion-based devices have been proposed to perform pattern recognition in ANNs  \cite{he2017tunable,song2020skyrmion}, skyrmionic synapses can potentially form more efficient neuromorphic hardware for SNNs.
State-of-the-art ANNs require convolutions which lead to massive matrix-vector multiplications.
Therefore, specialized hardware for ANNs focuses on making matrix operations faster and more power-efficient by using matrix acceleration units, e.g. Graphics Processing Unit (GPU) and Tensor Processing Unit (TPU).
In contrast, SNNs feature a massive connection of synapses and sparse activation of neurons.
Typically, the hardware implementations for SNNs entail the mapping of spiking neurons and their synapses into digital systems \cite{furber2014spinnaker,merolla2014million}, analog electronic circuits \cite{chicca2014neuromorphic}, and hybrid neuromorphic systems \cite{schemmel2010wafer} to achieve energy efficiency through event-driven computing.
Skyrmion-based neuromorphic hardware can be a promising candidate due to its potential for $i)$ storing the information in a non-volatile manner via the corresponding states of skyrmions, and $ii)$ energy-efficient device programming, due to their non-volatility and the promise of low current densities needed for manipulating the movement of skyrmions \cite{sampaio2013nucleation,fert2013skyrmions}.
Experiments in B20 systems have shown current densities as low as $10^{-4}$ MA/cm$^{2}$ \cite{jonietz2010spin,yu2012skyrmion,schulz2012emergent}.
More recent work on technologically relevant multilayers showed that high-speed ($\sim$ 100 m/s) manipulation of skyrmions in multilayer thin films required higher current densities ($\sim$10 MA/cm$^{2}$) \cite{woo2016observation,juge2019current,litzius2020role}.

Hitherto, all of the published numerical works on nanoscale skyrmion-based neuromorphic components, including skyrmion-based synapses \cite{huang2017magnetic,bhattacharya2019low} and skyrmion-based leaky-integrate-fire (LIF) neurons \cite{chen2018compact} were performed in ideal simulation conditions ($0$ K).
However, integrated systems, including SNNs, normally operate at room temperature (RT) which significantly deviates from the assumed (ideal) conditions of the previous works.
Using a zero temperature simulation of skyrmionic devices integrated into hybrid skyrmion-CMOS systems, including neuromorphic systems, can therefore be potentially misleading.
This paper aims to fill this important gap by investigating skyrmion-based neuromorphic computing components for stable RT operation with realistic parameters and device structures.

\begin{figure*}
\includegraphics[width=129mm]{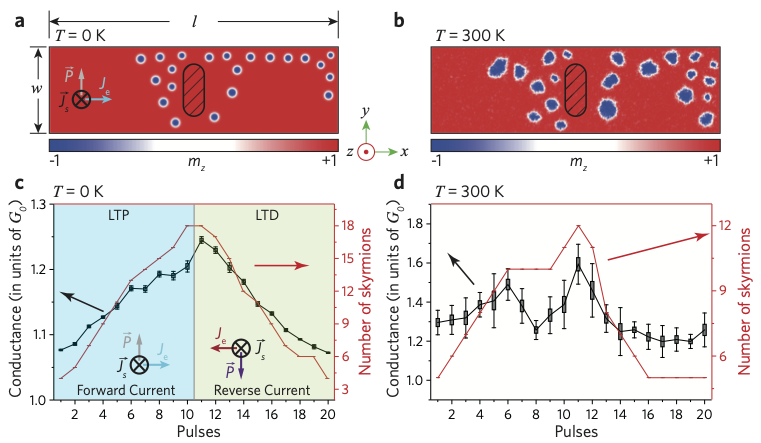}
\caption{\label{fig:2}The characteristic conductance modulation curves of the skyrmionic synapse.
Micromagnetic simulations of a skyrmionic synapse device at (a) $T = 0$ K, and (b) $T = 300$ K.
The length of the device is $l$ = $800$ nm and the width is $w$ = $220$ nm.
The patterned area indicates a rounded rectangle barrier with a size of $145$ nm $\times$ $60$ nm.
The color code (red-blue) is the same for (a) and (b).
The evolution of conductance (in units of $G_0$) (left $y$-axis) and the number of skyrmions (right $y$-axis) within the post-synapse region during the whole LTP/LTD process at (c) $T = 0\mathrm{K}$ and (d) $T = 300\mathrm{K}$.
Note that we apply $10$ electric current pulses in $+x$ direction and then $10$ electric current pulses in $-x$ direction, respectively.
The error-bars represent the standard deviation of each data point, obtained by $100$ distinct calculations of conductance of the same device (post-synapse region) at each state (defined by the intra-pulse period).}
\end{figure*}

We propose a RT skyrmionic synapse and investigate it systematically via micromagnetic simulations.
Instead of utilizing a simple structure of ferromagnetic metal ($\mathrm{FM}$) / heavy metal ($\mathrm{HM}$) \cite{huang2017magnetic,bhattacharya2019low}, we numerically design and evaluate a skyrmionic synapse with a tailored MML structure composed of repetitive $\mathrm{[HM_1/FM/HM_2]}$ sandwiched tri-layers inspired by \cite{moreau2016additive,boulle2016room}.
Micromagnetic simulations show that the operational stability of our proposed skyrmionic MML synapse is improved at RT ($300$ K).
The stability and robustness of the device can be further enhanced by modifying the number of repeated MML stacks and the structure of the device.
To improve the synaptic resolution of the skyrmionic synapse we use a stacked MMLs device structure with $4$ repeated MMLs.
This synapse can embed six discrete synaptic states with an estimated energy consumption of $\sim 300$ $\mathrm{fJ}$ per conductance state update event.

The proposed skyrmionic synapse is firstly integrated in an SNN framework and used for digit recognition exploiting the spike-timing-dependent plasticity (STDP) rule.
Using this unsupervised learning rule our skyrmionic SNN achieves $\sim 78\%$ classification accuracy on the MNIST handwritten digit data set, which is around $10\%$ lower than ideal synapses \cite{diehl2015unsupervised}.
To fully utilize the limited precision of synaptic weights and the intrinsic merits of skyrmionic synapses (non-volatility and energy-efficient synaptic programming), we then integrate the synapses into a deep SNN architecture - a feed-forward fully connected multilayer SNN - that has been shown to achieve higher performance levels than shallow SNNs \cite{tavanaei2019deep,diehl2015fast}.
Specifically, we employ a biologically inspired deep SNN according to Dale's principle \cite{dale1935pharmacology}.
The classification accuracy improves significantly ($\sim 98.61\%$) with only six synaptic states.
Furthermore, we propose a way to reboot the skyrmionic synapse after reaching equilibration by introducing an initialization programming pulse.
This pulse successfully recovers the synaptic modulation curve from a ``cold'' start.
The findings reported here provide further possibilities for the skyrmion-based neuromorphic computing.

\section{\label{sec:level1}Nanoscale room-temperature skyrmionic synapse in MMLs}
A synapse [Fig. \ref{fig:1}(c)] in the mammalian neocortex refers to a specialized junction that allows cell-to-cell communication.
It is widely accepted that the synapse plays a role in the formation of memory in the membrane brain \cite{dale1935pharmacology}.
Like the biological synapse, the schematic of a skyrmionic synapse is illustrated in Fig. \ref{fig:1}(d), composed of a pre-synapse region, a post-synapse region, and a barrier located in between \cite{huang2017magnetic,bhattacharya2019low}.
The barrier is one of the most important parts of this proposed synapse.
It can be achieved by locally tuning the anisotropy in the barrier area.
It is envisaged that this could be done by voltage-controlled magnetic anisotropy (VCMA) effect \cite{maruyama2009large} or ion irradiation \cite{chappert1998planar}.
Magnetic skyrmions can be nucleated in FM layers and driven in the track along the $x$-direction, as shown in Fig. \ref{fig:1}.
In such a synapse, synaptic weights are represented by the conductance of the post-synapse region measured from the magnetic tunnel junction (MTJ) reading device by applying an out-of-plane perpendicular reading current \cite{thomas2014perpendicular,zhang2018skyrmions}.
With appropriate current pulses applied (typically $\mathrm{MA/cm^2}$) skyrmions can be driven around the barrier.
The skyrmions are then trapped in the post-synapse region unless a current in the reverse direction is used.
The motion of skyrmions can be controlled by a current in-plane (CIP) flowing through the nanotrack or by a current perpendicular to the plane (CPP) \cite{tomasello2014strategy}.
When applying a CPP, a skyrmion receives a larger Slonczewski in-plane torque than the field-like out-of-plane torque generated by an equal CPP current density.
Skyrmions obtain higher velocities under CPP for a given current density.
Therefore, in this paper, we consider the case of CPP.

\begin{figure*}
\includegraphics[width=129mm]{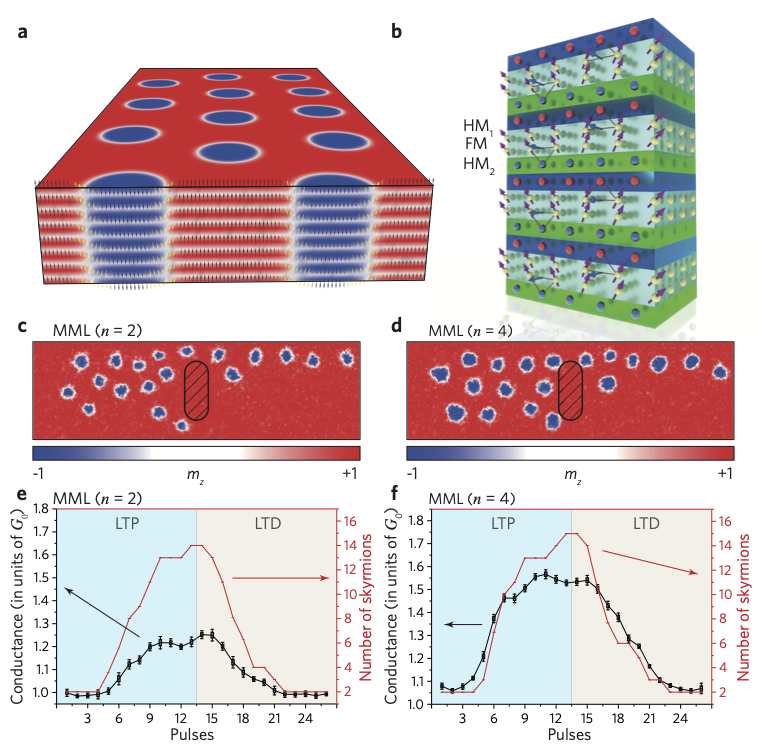}
\caption{\label{fig:3}The proposed skyrmionic synapses composed of the MML structure.
(a) Cross-sectional view of the simulation volume for a $\mathrm{[HM_1/FM/HM_2]_8}$ multilayer device where skyrmions assemble. Arrows point in the direction of the magnetization \textbf{\textit{m}}, and $\textbf{\textit{m}}_z$ is color-coded from blue ($-1$) to red ($+1$).
(b) Illustration of the MML structure comprised of several repetitions of the tri-layers, where ($\mathrm{FM}$, light blue) are sandwiched between two different heavy metals $\mathrm{HM_1}$ (blue, $\mathrm{Pt}$ in this paper) and $\mathrm{HM_2}$ (green, $\mathrm{Ir}$ in this paper) that induce a net enhanced DMI vector (where $\mathrm{HM_2}$ is the underlayer, and $\mathrm{HM_1}$ is the top layer, respectively).
Results of micromagnetic simulations of the proposed skyrmion-based synaptic devices with (c) MML ($n$ = $2$) and (d) MML ($n$ = $4$) in realistic RT conditions.
The color code (red-blue) is the same for (c) and (d).
Evolution of conductance and number of skyrmions for the post-synapse region with (e) MML ($n$ = $2$) and (f) MML ($n$ = $4$).}
\end{figure*}

The measured conductance depends upon the number of skyrmions in the post-synapse region underneath the MTJ reading device.
A magnetoresistance change is obtained, which is directly proportional to the number and size of skyrmions in the post-synapse area \cite{zhang2018skyrmions}.
Here, the FM layers at the reading region serve as the ``free layer'' of the MTJ device, while the top layer of the MTJ device (blue region in Fig. \ref{fig:1}(d)) serves as the ``fixed layer''.
The conductance is denoted as $G_{\mathrm{skr}}$ ($G_0$) with the presence (absence) of magnetic skyrmions.
The difference of two conductance states ($G_{\mathrm{skr}}$, $G_0$) can be controlled by tunneling magnetoresistance ratio (TMR).
In theory, the TMR ratio can be as large as $1,000\%$ in MTJs with MgO barriers \cite{mathon2001theory}.
In this paper, we assume $G_0 =$ $20$ $\mathrm{\mu S}$ and the TMR ratio to be $280\%$ as demonstrated experimentally \cite{parkin2004giant}.
Therefore, variation of the conductance of skyrmionic synapses is programmable by current injections, which is similar to biological synapses. 

Previous work on skyrmionic synapses utilized a single FM layer on an HM layer and investigated the long term potentiation (LTP) and long term depression (LTD) behavior of the device via simulations \cite{huang2017magnetic,bhattacharya2019low}.
We firstly apply this simple structure in the skyrmionic synapse and evaluate its functionality via the micromagnetic package $\mathrm{mumax^3}$ \cite{vansteenkiste2014design}.
In order to obtain the characteristic conductance modulation curve of the synapse device, $10$ current pulses with a $2$ $\mathrm{ns}$ duration between $5$ $\mathrm{ns}$ intervals are injected into the device, which will induce the increase of the conductance to form the LTP.
Then, another $10$ current pulses in the reverse direction are applied to reduce the number of skyrmions in the post-synapse region, resulting in the LTD.
As shown in Figs. \ref{fig:2}(a) and \ref{fig:2}(c), the skyrmionic synapse performs well at $0\mathrm{K}$.
With the forward and reverse direction of current, we received 11 distinct synaptic states (including the background state $G_0$), leading to a pattern recognition accuracy of $\sim 80\%$ in an SNN with $400$ excitatory neurons, which is $\sim 10\%$ lower than ideal synapses, in agreement with the results reported in \cite{bhattacharya2019low}.

\begin{figure*}
\includegraphics[width=129mm]{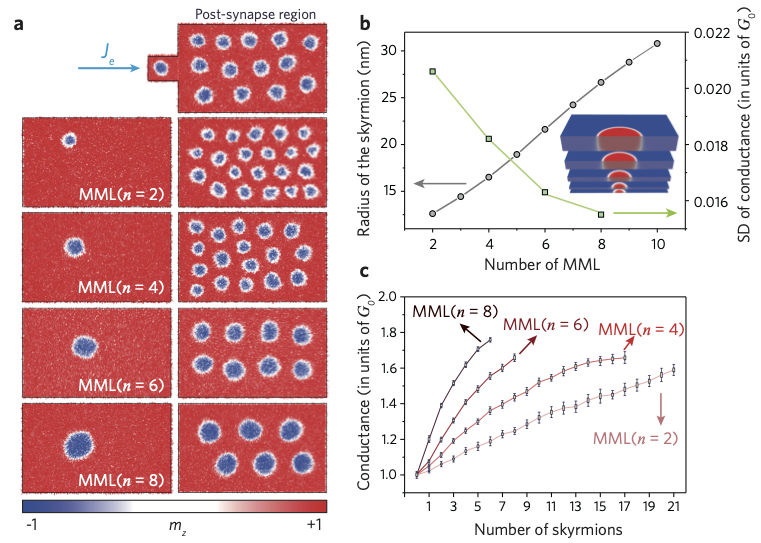}
\caption{\label{fig:4}Improvement in the device behavior as the number of MML tri-layers is changed.
(a) Micromagnetic simulations of the post-synapse region with $1$ skyrmion and full capacity of skyrmions in $2$, $4$, $6$ and $8$ repeated MMLs at RT.
$\mathrm{J_e}$ denotes the electric current applied within the HM layer in $+x$ direction.
(b) The radius of skyrmions and the standard deviation (SD) of calculated conductance with respect to the number of MML obtained from micromagnetic simulations. Inset: Schematics of a single skyrmion within MML systems with $n =$ $2$, $3$, $5$, $9$, and $15$ repeats, respectively.
(c) Conductance for the post-synapse region of the simulated synapse devices with varying MML structure as a function of the number of skyrmions hosted.}
\end{figure*}

In order to simulate the skyrmionic synapse at RT, we introduced finite-temperature effects through a randomly fluctuating thermal field \cite{brown1963thermal}:
\begin{equation}
    \vec B_{\mathrm{therm}} = \vec \eta(\textrm{step})\sqrt{\frac{2\alpha k_\mathrm{B} T}{\mu_0 M_{\mathrm{sat}}\gamma_{\mathrm{LL}}V\Delta t}},
\end{equation}
where $\alpha$ is the damping parameter, $k_\mathrm{B}$ the Boltzmann constant, $T$ the temperature, $M_{\mathrm{sat}}$ the saturation magnetization, $\gamma_{\mathrm{LL}}$ the gyromagnetic ratio, $V$ the cell volume, $\Delta t$ the time step and $\vec\eta(\textrm{step})$ a randomly oriented normal vector whose value is changed after each time step.

From the results in Figs. \ref{fig:2}(b) and \ref{fig:2}(d), only two distinct synaptic states can be roughly identified because of fluctuations of the shape of skyrmions.
The functionality of the skyrmionic synapse is catastrophically lost at $T$ = $300$K.
In order to stabilize RT skyrmions and have a working skyrmionic synapse that is technologically relevant, we propose a tailored MML [Fig. \ref{fig:3}(a)] structure in the skyrmionic synapse inspired by experimental observations \cite{moreau2016additive}.
The basic unit of the structure is a $\mathrm{[HM_1/FM/HM_2]}$ sandwiched tri-layer. The stabilization of skyrmions is enabled by the enhanced DMI from the asymmetric interfaces $\mathrm{[HM_1/FM]}$ and $\mathrm{[FM/HM_2]}$ \cite{moreau2016additive}, as shown in Fig. \ref{fig:3}(b).
Micromagnetic simulations with $\mathrm{[HM_1/FM/HM_2]}_n$ for $n = 2$ and $n = 4$ are shown in Figs. \ref{fig:3}(c) and \ref{fig:3}(d), respectively.
Compared to the simple $\mathrm{FM/HM}$ structure reported in previous work \cite{he2017tunable,bhattacharya2019low}, skyrmions in our proposed MML synapses exhibit enhanced thermal stability, which enables us to recover the characteristic conductance modulation curve of the RT skyrmionic synapses.
Figs. \ref{fig:3}(e) and \ref{fig:3}(f) depict the resulting synaptic weights (calculated conductance) with respect to the injected current pulses.
Skyrmionic synapses with MML ($n = 4$ repeats) exhibit a broader range of conductance as well as more distinct and discrete synaptic states than that with MML ($n = 2$).
Note that fewer synaptic states can be distinguished in skyrmionic synapses with MML ($n = 2$) due to the overlap and ambiguity among different states arising from thermal instabilities of skyrmions.

The improvement of performance metrics of the proposed MML skyrmionic synapse can be attributed to two primary reasons: $i)$ larger conductance ranges and $ii)$ more synaptic states, arising from the enhanced thermal robustness of the multilayer device.
As explained previously, the synaptic conductance variation range during the LTP/LTD process is calculated via the proportion of the ``free layer'' domain anti-parallel to the ``fixed layer'' domain, determined by the cross-sectional area and the number of skyrmions, so larger skyrmions enable a larger conductance range and greater ability to discriminate states.
We numerically constructed a series of half skyrmionic synapses (post-synapse region) where the number of repeated MMLs varies between $2$, $4$, $6$, and $8$, initially set to the background FM state.
Skyrmions are injected externally into the synapse one by one via current pulses, as shown in Fig. \ref{fig:4}(a).
Subsequently, the full capacity of the synapse and the stability of skyrmions are evaluated in these systems.
Due to the contribution of magnetostatic energy \cite{lemesh2018twisted}, as the number of MMLs increases, skyrmions with larger radii are stabilized in the system [Fig. \ref{fig:4}(b)].
Note that the standard deviation (SD) of the calculated conductance decreases in synapses with more repeated MMLs, indicating the better stability of skyrmions in the system and the thus the better stability of the skyrmionic synapse.
We calculated the conductance of each state and the results are shown in Fig. \ref{fig:4}(c).
Conductance ranges from $1.0G_0$ to $1.55G_0$ in $2$ MMLs by $11$ discrete states, $1.0G_0$ to $1.6G_0$ in $4$ MMLs by $14$ discrete states (the last 4 states are not distinguishable within the error bar), $1.0G_0$ to $1.65G_0$ in $6$ MMLs by $9$ discrete states, and $1.0$$G_0$ to $1.75$$G_0$ in $8$ MMLs by $6$ discrete states.
The skyrmionic synapse with $4$ MMLs has both relatively large synaptic weight range and largest number of distinguishable states.
Note that the results here (14 states) are different from Fig. \ref{fig:3}(f) (6 states), because in Fig. \ref{fig:3}(f), two skyrmions are injected simultaneously by each programming pulse, while here we artificially inject skyrmions one by one.

\begin{figure}
\includegraphics[width=86mm]{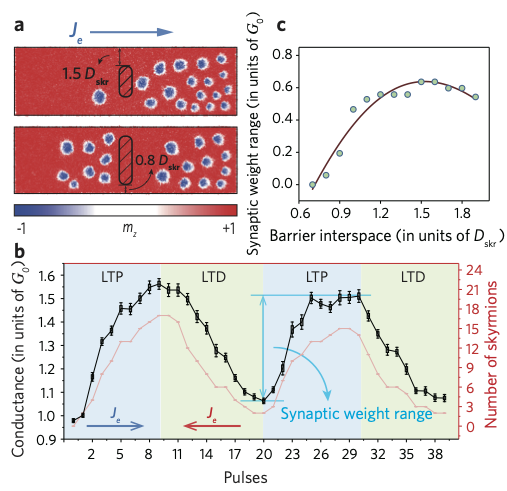}
\caption{\label{fig:5}Tuning the size of the barrier to improve the characteristic conductance modulation curves of skyrmionic synapses.
(a) Micromagnetic simulations of the skyrmionic synapse with the barrier interspace of $1.5$ $D_{\mathrm{skr}}$ and $0.8$ $D_{\mathrm{skr}}$ after the $10^{th}$ current pulse in the $x$ direction. (b) Evolution of conductance and number of skyrmions of the synaptic device with the barrier interspace of $1.2$ $D_{\mathrm{skr}}$ during two full LTP/LTD programming operations. (c) Synaptic weight range as a function of the interspace between the barrier and edges.}
\end{figure}

\begin{figure}
\includegraphics[width=86mm]{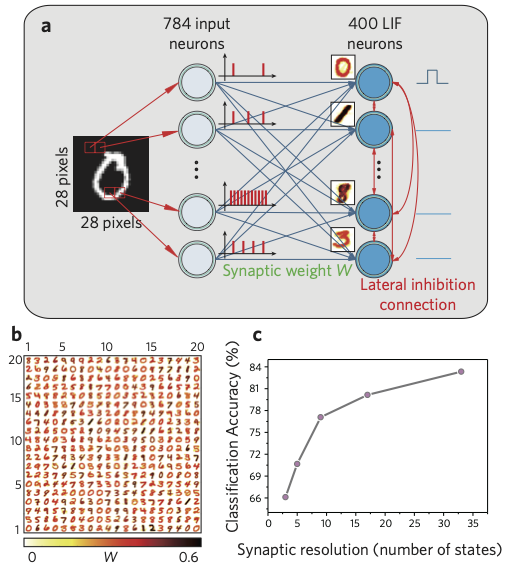}
\caption{\label{fig:6}The 2-layer skyrmionic SNN for unsupervised learning via STDP method.
(a) SNN structure for pattern recognition.
(b) Visual representation of the synaptic weights after the training process in SNNs of $400$ neurons with six discrete synaptic states (corresponding to the synaptic resolution achievable by the MML ($n$ = $4$) skyrmionic synapse).
Dark regions: higher weight values, indicating the pattern learned by the corresponding LIF post-neuron.
(c) Classification accuracy of SNN on the MNIST data set as a function of the number of distinct synaptic states.}
\end{figure}

\begin{figure}
\includegraphics[width=86mm]{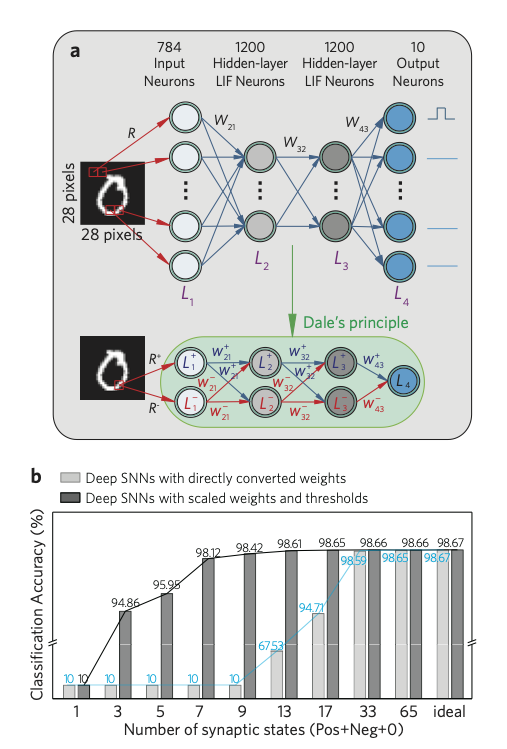}
\caption{\label{fig:7}The supervised skyrmionic deep SNN.
(a) Schematics of the proposed biologically inspired structure of the deep skyrmionic SNN utilizing Dale's principle. (b) Comparison of classification accuracy between i) skyrmionic deep SNNs with directly converted weights and ii) skyrmionic deep SNNs with scaled weights and thresholds, for different number of synaptic states.}
\end{figure}

\begin{figure}
\includegraphics[width=86mm]{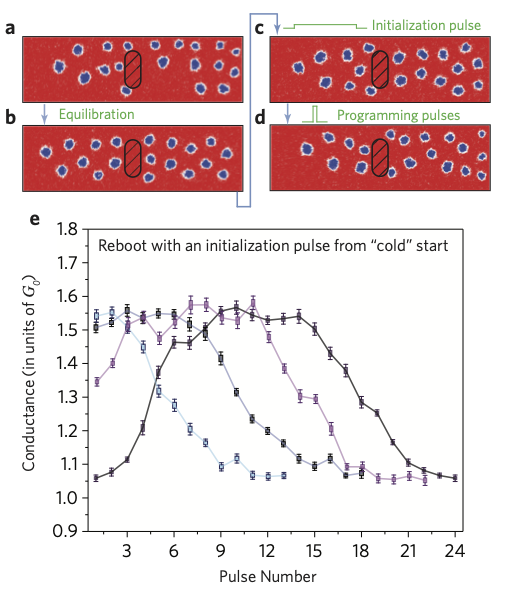}
\caption{\label{fig:8}Introducing an initialization pulse to reboot the skyrmionic synapse after equilibration.
Micromagnetic simulations for the MML ($n$ = $4$) skyrmionic synapse (a) after the $5^{th}$ pulse, and then (b) after the equilibration process. The skyrmionic synapse could be rebooted (c) through an initialization pulse and then (d) programmed via programming pulses. (e) Evolution of conductance modulation curves for the post-synapse region of the skyrmionic synapse starting from different synaptic states with an initialization pulse.}
\end{figure}

Interestingly, we also observe that the conductance modulation curves do not fully recover to the initial state after the first LTP/LTD process [Fig. \ref{fig:5}(a, b)].
This situation is due to a few skyrmions being constrained in the pre-synapse (post-synapse) region no matter how many current pulses in the $+x$ ($-x$) direction are applied.
The number of remaining skyrmions increases with larger barriers, as shown in Figs. \ref{fig:5}(a).
However, if we take the first LTP/LTD operation as a calibration process [Fig. \ref{fig:5}(b)], the conductance range stays unchanged during subsequent programming operation.
Therefore, we define the valid synaptic weight range of the RT skyrmionic synapse as the conductance range during the second LTP/LTD process.
In order to explore further possibilities for tuning of the synaptic resolution, we carefully modified the interspace between the barrier and edges from $0.7D_{\mathrm{skr}}$ to $2.0D_{\mathrm{skr}}$ where $D_{\mathrm{skr}}$ represents the diameter of a single skyrmion in the synapse from Fig. \ref{fig:4}(b). 
The calculated synaptic weight range with respect to the barrier interspace is shown in Fig. \ref{fig:5}(c).
We find that the skyrmions in the post-synapse region will cross the barrier back to the pre-synapse region during equilibration when the barrier interspace is larger than $1.7D_{\mathrm{skr}}$, resulting in the downward trend of the second half of the curve in Fig. \ref{fig:5}(c). 
A peak value of the synaptic weight range of $0.6G_0$ at barrier interspace of $1.5D_{\mathrm{skr}}$ is obtained from the Gaussian fitting of the discrete data points.
Therefore, the skyrmionic synapses we mainly discuss in this paper are based on the $4$ MML structure and $1.5D_{\mathrm{skr}}$ barrier interspace unless specified otherwise, including results of Figs. \ref{fig:2} and \ref{fig:3}.
Thermal stability can be optimized by tuning the multilayer stack, the original nucleated skyrmion density, the device size and the driving current amplitude, to ensure that the device can operate without significant skyrmion collapse.

The energy consumption per synaptic update event is given by \cite{bhattacharya2019low}: 
\begin{equation}
    E_{\mathrm{update}} = 2\rho tlwJ^2T_{\mathrm{pulse}},\label{CalculatingEnergy}
\end{equation}
where $\rho$ is the resistivity of $\mathrm{Pt}$ thin film \cite{nguyen2016spin}, $l$, $t$ and $w$ are the length, thickness and width of the top and bottom $\mathrm{HM}$ layers, $J$ the amplitude of current density, and $T_{\mathrm{pulse}}$ is the duration of the pulses applied on the device, so the total thickness is $2t$.
The programming energy to update the synaptic state $E_{\mathrm{update}}$ in the proposed synapse is estimated $\sim 300$ $\mathrm{fJ}$ from Eq. (\ref{CalculatingEnergy}) by using the simulation parameters in Appendix \ref{AppendixA}.

The proposed multilayer skyrmionic synaptic device is experimentally feasible with industrially relevant material systems at lithographically accessible length scales. Furthermore, skyrmion-based devices can be technically compatible with CMOS circuits in hybrid spintronic-CMOS systems such as those proposed in \cite{bhattacharya2019low,sengupta2016hybrid}, 
which may enable further integration of skyrmionic synapses in neuromorphic hardware for pattern recognition and dynamic signal analysis tasks.

\section{\label{sec:level1}SKYRMIONIC SNNS FOR PATTERN RECOGNITION}
To validate the functionality of the proposed RT skyrmionic synapses, we simulate a 2-layer unsupervised SNN.
We then propose that skyrmion-based systems can also be deployed within a supervised deep SNN in order to achieve superior accuracy in pattern recognition tasks.
The input synaptic resolution (six synaptic states) and the synaptic range ($0.6G_0$) of the nanoscale skyrmionic synapse are derived from micromagnetic simulations at RT [Fig. \ref{fig:3}(f)].
The detailed illustration of the simulation set-ups can be found in Figs. \ref{fig:6} and \ref{fig:7}, and Appendix \ref{AppendixB}.
The proposed skyrmionic synapses can be embedded into a crossbar array \cite{chakraborty2018technology} to connect adjacent layers of neurons and to provide synaptic weights in SNNs.
The LIF neurons can be implemented by analog circuits demonstrated in \cite{chicca2014neuromorphic}.
The crossbar hardware and LIF neurons could be implemented according to \cite{chakraborty2018technology,payvand2019neuromorphic}, but that is beyond the scope of the present work.

We first simulate a fully-connected 2-layer SNN inspired from \cite{diehl2015unsupervised}, as shown in Fig. \ref{fig:6}(a).
The STDP rule, which states that the synaptic weights should increase (decrease) when the pre-neuron fires earlier (later) than the post-neuron \cite{brader2007learning,serrano2013stdp}, trains the SNN on $60,000$  images of the MNIST handwritten data set using BRIAN, a Python-based simulator for neuromorphic computing \cite{goodman2008brian}.
The MNIST data set consists of 10 classes (digits 0 $\to$ 9) on a grid of 28 $\times$ 28 pixels, which is widely used as a standard benchmark.
Note that the proposed skyrmionic SNNs will be mainly suitable for practical application scenarios that deal with dynamic signals, such as sensory signal processing, brain-machine interfaces, robot control, etc.
The input-layer neurons encode these $28\times28$ pixels images as Poisson spike trains with an average firing rate proportionate to the intensity of pixels.
The $784$ input neurons and $400$ LIF neurons are fully connected, while the lateral inhibition connection between $400$ LIF neurons is one-to-all.

Fig. \ref{fig:6}(b) shows the rearranged weights (from 784 to 28 $\times$ 28) of the connections from input to excitatory neurons in a 20 $\times$ 20 grid. The visually displayed patterns correspond to the response digit classes of $400$ LIF neurons.
Although there are only 10 classes of patterns in the training set, larger number of excitatory LIF neurons will correspond to higher network performance, because multiple neurons are assigned to each digit class after training \cite{diehl2015unsupervised}.
The classification accuracy of the proposed skyrmionic SNN after training is $\sim 76\%$ and converges after around $100$ training time steps.
The trained SNN is evaluated through $10,000$ MNIST images, where an accuracy of $\sim 78\%$ is obtained, which is $\sim 10\%$ lower than ideal synapses illustrated in \cite{diehl2015unsupervised}.
The result shown in Fig. \ref{fig:6}(c) indicates that the large number of synaptic states is highly desirable for high classification accuracy.
However, more skyrmionic synaptic states require larger devices and more programming energy, which will counteract the advantages of our proposed RT synapse.

To gain a better performance of the skyrmionic SNNs, we then design and evaluate a deep SNN with proposed RT skyrmionic synapses.
Although high-precision synaptic weights are essential to obtain converging results during the training process, neural networks are able to operate with limited-precision weights during the inference process with acceptable accuracy loss \cite{stromatias2015robustness}, which is compatible with limited precision and low power hardware platforms as we proposed in this work.

Here, we demonstrate a prototype and basic operations of skyrmionic synapses in a simple deep SNN structure, which can also be further upgraded to other state-of-the-art deep neural networks \cite{lecun2015deep}.
The deep SNN we investigate consists of an input layer, an output layer, and two hidden layers.
The $784$ input neurons, $2$ hidden layers with $2,400$ neurons per layer, and $10$ output neurons are fully connected in sequence.
In order to deploy RT skyrmionic synapses in deep SNNs, we propose a biology-inspired structure of SNNs, according to Dale's principle, by doubling the number of hidden layer neurons and splitting synaptic weights to positive $W^+$ and negative value $W^-$ as shown in Fig. \ref{fig:7}(a).
At the same time, the direct conversion of synaptic weights from full to limited precision may cause significant accuracy drop in SNNs \cite{stromatias2015robustness,mahajan2018exploring}.
Therefore, we also propose a conversion method in R.T skyrmionic SNNs: introducing a scale factor $\sigma$, which significantly increases the accuracy for SNNs with low-precision weights.
More details for training and conversion of the skyrmionic deep SNNs are provided in Appendix \ref{AppendixB}.

We simulate the deep SNN with different precision of weights by changing the number of synaptic states to $1$, $3$, $5$, $7$, $9$, $13$, $17$, $33$, $65$, and $+\infty$.
The number of synaptic states is given by $2X + 1$, where $X$ is the number of positive and of negative synaptic weights, and there is one zero-weight state.
The negative values can be obtained by applying a reverse voltage in the crossbar hardware implementations \cite{chakraborty2018technology}, and the zero-weight state can be acquired by setting the same value of $W^+$ and $W^-$ in Fig. \ref{fig:7}(a).
For example, the number of $13$ synaptic states consists of $6$ positive, $6$ negative and a zero-weight state.
The classification accuracy for each weight precision is illustrated in Fig. \ref{fig:7}(b), where the height difference between the light grey and dark grey columns represents the improvement in accuracy of SNNs with directly converted synaptic weights to SNNs with scaled weights and thresholds.
For SNNs with directly converted weights, the results show that the skyrmionic synapse should have $33$ synaptic states ($16$ skyrmionic states) to achieve a $<1\%$ accuracy loss compared to the ideal full-precision synapses.
In comparison, SNNs with scaled weights and thresholds show a much faster increase of classification accuracy when the number of synaptic states increases.
The accuracy exceeds $\sim 98\%$ at only $7$ synaptic states ($3$ skyrmionic synaptic states).
Notably, we obtain a superior $\sim 98.61\%$ classification accuracy with $13$ synaptic states ($6$ skyrmionic states of RT skyrmionic synapse), which is merely $\sim 0.06\%$ lower than the SNNs with ideal full-precision synapses.
The results here demonstrate the excellent potential for the use of the proposed skyrmionic synapses in neuromorphic computing, especially when deployed in deep SNNs and ensuring RT operation.

\section{\label{sec:level1}DISCUSSION}
In the micromagnetic simulations the motion of skyrmions is induced by a series of CPP pulses with a fixed pulse duration.
We observed that, over time, skyrmions encounter greater difficulty to cross the barrier.
With the majority of skyrmions passing over the barrier, fewer skyrmions are left in the pre-synapse region.
Consequently, skyrmion-skyrmion repulsion is less likely to enable the crossing of the barrier, also highlighted in \cite{bhattacharya2019low}.
Once there is a long enough time interval between adjacent programming operations, skyrmions equilibrate into a uniform distribution, resulting in a latency and even failure for updating synaptic weights.

In order to address this problem we propose a way to operate the device from a ``cold'' start.
We introduce a small-amplitude initialization pulse before any programming pulses.
In the simulations, we set a current pulse with a duration of $25$ $\mathrm{ns}$ and an amplitude of $5$ $\mathrm{MA/cm^2}$ ($1/10$ of the programming pulses), as shown in Fig. \ref{fig:8}.
After the implementation of the initialization pulse, skyrmions in Fig. \ref{fig:8}(c) form a similar ensemble as during the programming process in Fig. \ref{fig:8}(a).
Simulations demonstrate that this method ensures the robust operation of the skyrmionic synapse, as it can be rebooted from a ``cold'' start to recover the modulation curve from an arbitrary initial state, as shown in Fig. \ref{fig:8}(e).

Compared to CMOS-modelled memristors \cite{pershin2010experimental} and floating-gate memory cells \cite{hasler2013finding}, the proposed skyrmionic devices utilize the evolution/propagation of the magnetic excitations as information carriers rather than the movement of electrons/holes themselves.
This different nature of information processing offers opportunities for energy-efficient computing and storage by harnessing topological twists in the magnetic fabric.
Moreover, skyrmionic synapses have advantages of shorter update times and relatively smaller cycle-to-cycle and device-to-device variability compared to existing transistor-free phase change memory (PCM) devices \cite{boybat2018neuromorphic} and resistive random access memory (RRAM) \cite{prezioso2015training}.
Therefore, the skyrmionic devices are promising for applications in low-power neuromorphic computing and inference tasks in edge computing devices \cite{han2019convergence} (see Appendix \ref{AppendixC} for a proposed scheme).
Moreover, further research should be undertaken to investigate the high accuracy and low power $in$ $situ$ training of skyrmionic SNNs utilizing the on-chip surrogate gradient technique \cite{neftci2019surrogate} and the Federated Learning (FL) for collaborative inter-device learning \cite{skatchkovsky2019federated}.

\section{\label{sec:level1}Conclusion}
In order to enable the potential of skyrmionic devices for neuromorphic computing, we demonstrate the stabilization of magnetic skyrmions at RT in MMLs tailored for improving synaptic resolution.
In this work, we propose a nanoscale multilayer skyrmionic synapse for deep SNNs.
Instead of the single $\mathrm{FM/HM}$ device structure utilized in existing work, we propose a tailored MML structure with repeated sandwiched stack $\mathrm{[HM_1/FM/HM_2]}$, which enhances the skyrmions' stability in order to ensure robust RT synaptic functionality and enable integration in an SNN framework.
We use the number of MMLs repeats and the size of the barrier to tune the thermal stability of skyrmions and the desired synaptic profile and resolution.
We then embed the skyrmionic synapses into SNNs to demonstrate the pattern recognition.
Firstly a 2-layer skyrmionic SNN is established and trained by the unsupervised STDP method.
The functionality of the network is evaluated utilizing the MNIST handwritten digit data set.
We obtain a classification accuracy of $\sim 78\%$ and approximately $10\%$ drop of accuracy from ideal synapses with full-precision weights.
To fully exploit the limited-precision weights and the intrinsic merits of RT MML skyrmionic synapses, we integrate the skyrmionic synapse into a deep SNN.
A high classification accuracy of $\sim 98.61\%$ is achieved with six skyrmionic synaptic states obtained from the proposed RT MML ($n$ = $4$) skyrmionic synapses.
The emulation of deep SNNs with our proposed RT skyrmionic synapses enables wider possibilities for energy-efficient hardware implementations to perform neuromorphic computing.

\begin{acknowledgments}
RC and YL wish to acknowledge the China Scholarship Council (CSC) and the University of Manchester for the funding support.
\end{acknowledgments}

\appendix
\section{MICROMAGNETIC SIMULATIONS}
\label{AppendixA}
The micromagnetic simulations were performed using the GPU-accelerated micromagnetic programme $mumax^3$.
The time-dependent magnetization dynamics are described by the Landau-Lifshitz-Gilbert (LLG) equation. 
\begin{equation}
    \frac{d\textbf{\textit{m}}}{dt}=-|\gamma_{\mathrm{LL}}|\textbf{\textit{m}}\times \textbf{\textit{h}}_{eff}
    +\alpha \textbf{\textit{m}} \times \frac{d\textbf{\textit{m}}}{dt}
    +\frac{u}{t}\textbf{\textit{m}} \times (\textbf{\textit{m}}_\mathrm{p} \times \textbf{\textit{m}}),
\end{equation}
where $\textbf{\textit{m}}=\textbf{\textit{M}}/M_s$ is the reduced magnetization,
$M_s$ is the saturation magnetization, $\gamma_{\mathrm{LL}}$ is the gyromagnetic ratio,
$\textit{\textbf{h}}_{eff}=H_{eff}/M_s$ is the reduced effective field, $\alpha$ is the damping parameter, $t$ is the thickness of the $\mathrm{FM}$ layer, and $\textit{\textbf{m}}_\mathrm{p}$ is the polarization direction.
The energy density contains the exchange energy term, the anisotropy energy term, the Zeeman energy term, the magnetostatic energy term and the DMI energy term.
We consider a tailored sandwich structure of $\mathrm{[HM_1/FM/HM_2]}_n$ with perpendicular magnetic anisotropy and interfacial DMI. The input material parameters to perform the simulations are chosen according to the reported experimental results \cite{moreau2016additive}.
Damping parameter $\alpha$ = 0.1, DMI constant $D_{\mathrm{ins}} = \mathrm{1.7~mJ\cdot m^{-2}}$, Gilbert gyromagnetic ratio $\gamma_{\mathrm{LL}} = -2.211 \times\mathrm{10^5~mA^{-1}\cdot s^{-1}}$, saturation magnetization $M_{s} = \mathrm{956~kA\cdot m^{-1}}$, the uniaxial out-of-plane magnetic anisotropy $K_u = \mathrm{717~kJ\cdot m^{-3}}$, and the exchange stiffness $A = \mathrm{10~pJ\cdot m^{-1}}$.
The spin Hall polarization $\Theta_{SH}$ is chosen as 0.6 following Refs. \cite{bhattacharya2019low,huang2017magnetic}.
An external magnetic field of $80$ $\mathrm{mT}$ in the out-of-plane direction is applied. A higher magnetic anisotropy $K_{\mathrm{u,high}} = \mathrm{860~kJ\cdot m^{-3}}$ is set for the barrier of the device. The size of the whole device is $800$ nm $\times$ $220$ nm $\times$ $3n$ nm (where $n$ denotes the number of repeated MML). In order to ensure the accuracy of calculation, the mesh size is set to $2$ nm $\times$ $2$ nm $\times$ $1$ nm, which is less than the exchange length $l_{\mathrm{ex}} = 2\sqrt{A/(\mu_0M_s^2)} = 5.9~\mathrm{nm}$ and DMI length $l_{DMI} = 2A/D = 11.77~\mathrm{nm}$. In our multilayers, the intermediate $\mathrm{HM_1}$ and $\mathrm{HM_2}$ layers are thinner than the spin diffusion length.
In this case the torques would be efficient only in the external layers \cite{legrand2018hybrid}.
In this work the spin orbit torque (SOT) created via CPP is applied only in the first bottom and the first top layers and the injected spin polarization is uniform in these two layers.
The injected current is then modeled as a fully polarized (along $+y$ direction) vertical spin current of current density $J$ = $30$ $\mathrm{MA\cdot cm^{-2}}$.

\section{SNN SIMULATIONS ON THE MNIST DATA SET}
\label{AppendixB}

\begin{figure}
\includegraphics[width=86mm]{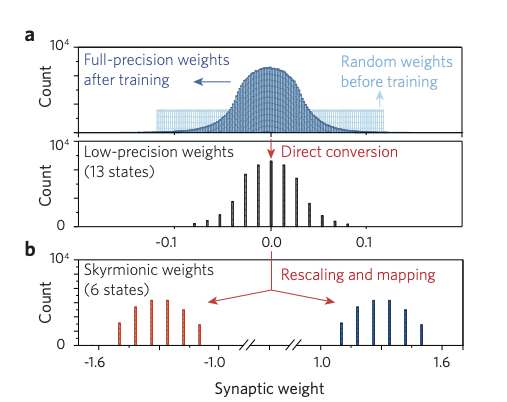}
\caption{\label{fig:9}Conversion and training of skyrmionic deep SNNs.
(a) Weight distributions for the full-precision weights in the 4-layer ANN before and after training, and directly converted low-precision weights with 13 discrete states.
(b) By using Dale's principle, we re-scale and map the weights into the deep SNN with six discrete synaptic states obtained at RT by the MML ($n$ = $4$) skyrmionic synapse.
The negative value marked as red columns in the left can be obtained by applying a reverse voltage in the crossbar hardware implementations.
}
\end{figure}

For the simulation of the unsupervised skyrmionic SNN, we utilized Python and the BRIAN simulator \cite{goodman2008brian}.
For the MNIST pattern recognition, we simulated a 2-layer SNN ($784$ neurons as input layer and $400$ neurons as excitatory layers) on BRIAN inspired by \cite{diehl2015unsupervised}.
The proposed skyrmionic SNN updates synaptic weights by a discrete STDP process adapted from \cite{diehl2015unsupervised}, which means the weights are modified to the adjacent synaptic value within the six synaptic states in Fig. \ref{fig:3}(f) during training iterations. The skyrmionic SNN is trained with $60,000$ images of MNIST handwritten data set. The trained SNN is tested with another $10,000$ images.

With respect to the simulation of the supervised skyrmionic deep SNN, the proposed network is obtained through the technique of ANN-to-SNN conversion, which consists of three steps:
$i)$ We first train a 4-layer ANN with full-precision weights. 
$ii)$ We then directly convert the weights of the trained ANN to low-precision synaptic weights with 13 states [Fig. \ref{fig:9}(a)].
$iii)$ Finally, in order to map the weights into six discrete synaptic states obtained at RT by the MML ($n$ = $4$) skyrmionic synapse, we re-scale the synaptic weights via tuning the hyper-parameters of SNNs (e.g. neuron thresholds).

We utilize the DeepLearn Toolbox on Matlab for the training of the ANN, where layers are fully connected by the weight matrices $W_{21}$, $W_{32}$, and $W_{43}$, respectively [Fig. \ref{fig:7}(a)].
The value of weight matrices is initialized randomly between $-0.1$ and $0.1$ [upper Fig. \ref{fig:9}(a)].
We choose the Rectified Linear Unit (ReLU) activation function and the stochastic gradient descent (SGD) training method, where the batch size is $100$ and the number of training epochs is $15$. 
Simulation of the SNN is based on the method presented in \cite{diehl2015fast}, implemented in Matlab to enable us to introduce noise and modify weight matrices.
The input encoding method is rate coding, where the input firing rate is proportional to the intensity of pixels.
We set the maximum input firing rate of SNNs to be $1$ $\mathrm{kHz}$ and the time resolution to be $1$ $\mathrm{ms}$.
The neuron model is LIF with leaky time constant of $50$ $\mathrm{ms}$. The threshold of neurons is initially set to $1$.

The limited precision and the discrete space of the skyrmionic synapse may hinder its applications of building high-accuracy SNNs.
We propose here a conversion method and constructing SNNs according to Dale's principle, which significantly increases the accuracy with limited precision weights.
Supervised learning of neural networks requires the weights to be axisymmetric with zero point, which is yet satisfied by skyrmionic synapses.
We address the issue by amending the structure of the SNN as follows.
The original weight matrices of the SNN are $W_{21}$, $W_{32}$, and $W_{43}$ which connect the four layers $L_1$, $L_2$, $L_3$, and $L_4$ of the SNN in Fig. \ref{fig:7}(a).
The input is represented as $R$ and is fed into $L_1$ to determine whether neurons fire or not.
In the structure shown in the lower part of Fig. \ref{fig:7}(a), the number of neurons in the input layer and two hidden layers is doubled, which forms the neurons $L_1^+$, $L_1^-$, $L_2^+$, $L_2^-$, $L_3^+$, and $L_3^-$.
Meanwhile, the neurons in the output layer keep unchanged as $L_4$.
The inputs $R^+$ and $R^-$ are fed into $L_1^+$ and $L_1^-$, respectively, with the same value as R.
The weight matrices $W_{21}$, $W_{32}$, and $W_{43}$ are split to positive weight matrices $W_{21}^+$, $W_{32}^+$, and $W_{43}^+$ and negative weight matrices $W_{21}^-$, $W_{32}^-$, and $W_{43}^-$ with identical size.
The value of weights is within the range of skyrmionic synaptic weights $\{[-1.6,-1.0], [1.0,1.6]\}$, as shown in Fig. \ref{fig:9}(b).

The proposed rescaling method [Fig. \ref{fig:9}(b)] comes from the idea that the accuracy loss before and after conversion is due to the mismatch of synaptic weights.
For example, most weighs are distributed in the range between $-0.1$ and $0.1$ after training of the ANN, while the synaptic states obtained at RT
by the MML ($n$ = $4$) skyrmionic synapse range between $-0.6$ and $0.6$.
This mismatch could be eliminated by scaling the directly converted low-precision weights to a proper weight range, where the stored critical information could be adequately represented in low-precision synapses.
Therefore, a scaled factor $\sigma = 0.6/0.1 = 6$ is applied on the thresholds of neurons during the conversion.
In our approach, the scale factor is obtained by scanning $\sigma$ and finding the value of the scaled factor achieving the highest accuracy.
Note that the dark grey columns in Fig. \ref{fig:7}(b) show the classification accuracy with the appropriate scaled factor $\sigma$ applied on the thresholds of neurons in each layer.

\section{SKYRMIONIC DEEP SNN FOR EDGE COMPUTING}
\label{AppendixC}

\begin{figure}
\includegraphics[width=86mm]{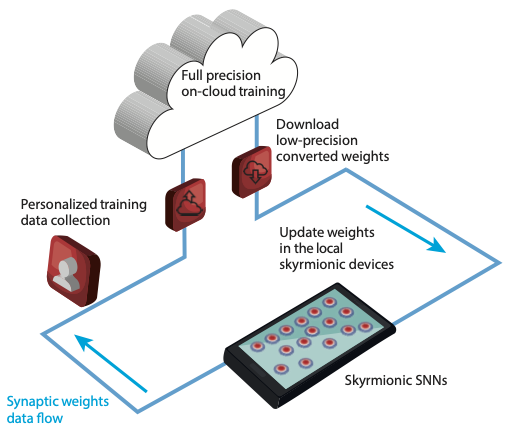}
\caption{\label{fig:10}Skyrmionic deep SNNs for edge computing.
Schematic shows a possible scenario of use. The proposed skyrmionic deep SNN can be integrated into a framework concept where training takes place in the cloud and updating at the edge. The process is an iterative full-precision training to low-precision conversion cycle: $i)$ full precision cloud-based online training and $ii)$ skyrmionic devices low-energy updating with offloaded low-precision synaptic weights.}
\end{figure}

The non-volatility, nanoscale footprint, energy-efficiency of the proposed skyrmionic synapse and the ability to operate at RT make it suitable for intelligent edge devices that can perform low-power and accurate neural networks inference such as pattern recognition and speech recognition.
In edge devices, the power supply can be limited. Therefore, only neural networks with high energy-efficiency both on the programming and internet of things (IoT) inferring aspects can be supported.

Here we propose a scenario of possible use of the skyrmionic synapses in an intelligent edge terminal, e.g. a smart wrist band to monitor the body blood pressure deployed in a deep SNN framework.
The training is accomplished using the standard SGD back-propagation method with full precision weights.
After the cloud-based online training finishes we transfer the full precision weights into low precision synaptic states and program the skyrmionic device at the edge by applying current pulses from the crossbar hardware to skyrmion-based synapses.
As different personalized training is necessary among individuals, the SNN can be retrained by fine-tuning starting from the merged personalized data set collected from the specific user.
The skyrmionic SNN is then sparsely updated with the downloaded synaptic weights with low energy consumption, which forms a closed-loop system, as depicted in Fig. \ref{fig:10}.

\nocite{*}
\bibliography{skyrmionicSynapse}

\end{document}